\documentstyle[11pt,aaspp4]{article}

\begin{document}

\null

\vskip 0.5truein

\title{An X-ray and Optical Investigation of the Environments\\
Around Nearby Radio Galaxies}

\author{Neal A. Miller\altaffilmark{1,2}} 
\authoremail{nmiller@aoc.nrao.edu}

\author{Frazer N. Owen\altaffilmark{1}}
\authoremail{fowen@aoc.nrao.edu}

\author{Jack O. Burns\altaffilmark{3}}
\authoremail{Burnsj@missouri.edu}

\author{Michael J. Ledlow\altaffilmark{4}}
\authoremail{mledlow@wombat.phys.unm.edu}

\and

\author{Wolfgang Voges\altaffilmark{5}}
\authoremail{whv@mpe.dnet.nasa.gov}

\altaffiltext{1}{National Radio Astronomy Observatory, P.O. Box 0, Socorro, NM 87801. The National Radio Astronomy Observatory is a facility of the National Science Foundation operated under cooperative agreement by Associated Universities, Inc.}

\altaffiltext{2}{New Mexico State University, Department of Astronomy, Box
30001/Dept. 4500, Las Cruces, New Mexico 88003}

\altaffiltext{3}{Office of Research and Department of Physics and Astronomy, University of Missouri, Columbia, MO  65211}

\altaffiltext{4}{Department of Physics and Astronomy, University of New Mexico, Albuquerque, NM  87131}

\altaffiltext{5}{Max-Planck-Institut f\"ur Extraterrestrische Physik, D-85740 Garching bei M\"unchen, Germany}

\newpage

\begin{abstract}
Investigations of the cluster environment of radio sources have not shown a correlation between radio power and degree of clustering. However, it has been demonstrated that extended X-ray luminosity and galaxy clustering do exhibit a positive correlation. This study investigates a complete sample of 25 nearby ($z \leq 0.06$) radio galaxies which are not cataloged members of Abell clusters. The environment of these radio galaxies is studied in both the X-ray and the optical by means of the {\it ROSAT All-Sky Survey} (RASS), {\it ROSAT} pointed observations, and the Palomar optical Digitized Sky Survey (DSS). X-ray luminosities and extents are determined from the RASS, and the DSS is used to quantify the degree of clustering via the spatial two-point correlation coefficient, $B_{gg}$. Of the 25 sources, 20 are $\ge 3 \sigma$ detections in the X-ray and 11 possessed $B_{gg}$'s significantly in excess of that expected for an isolated galaxy. Adding the criterion that the X-ray emission be resolved, 10 of the radio galaxies do appear to reside in poor clusters with extended X-ray emission suggestive of the presence of an intracluster medium. Eight of these galaxies also possess high spatial correlation coefficients. Taken together, these data suggest that the radio galaxies reside in a low richness extension of the Abell clusters. The unresolved X-ray emission from the other galaxies is most likely associated with AGN phenomena. Furthermore, although the sample size is small, it appears that the environments of FR I and FR II sources differ. FR I's tend to be more frequently associated with extended X-ray emission (10 of 18), whereas FR II's are typically point sources or non-detections in the X-ray (none of the 7 sources exhibit extended X-ray emission).
\end{abstract}

\section{Introduction}

Technological advances within the past twenty or so years have led to a greater understanding of the cluster environment. Sensitive radio telescopes such as the {\it Very Large Array} (VLA) have mapped radio emission originating from individual galaxies, with such emission frequently extending well beyond the physical size of the host galaxy ($\geq 100~h_{75}^{-1}$kpc). In addition, X-ray satellites such as {\it Einstein} and {\it ROSAT} have probed the hot X-ray emitting gas associated with galaxy clusters. In fact, X-ray observations are now frequently used to identify clusters and compact groups (see, e.g., \markcite{craw1996}Crawford \& Fabian 1996,\markcite{mulc1996} Mulchaey {\it et al.} 1996,\markcite{ostr1995} Ostriker {\it et al.} 1995,\markcite{pild1995} Pildis {\it et al.} 1995,\markcite{doee1995} Doe {\it et al.} 1995, \markcite{burn1996} Burns {\it et al.} 1996).

Radio studies have not shown a correlation between the probability that a galaxy is a radio source and the degree of local clustering (see, e.g., \markcite{leda1995}Ledlow \& Owen 1995). Specifically, the radio luminosity function does not differ for cluster vs. field radio galaxies (\markcite{fant1984}Fanti 1984, \markcite{ledl1996}Ledlow \& Owen 1996), nor does the probability for a galaxy to exhibit radio emission depend on global cluster galaxy density (e.g., \markcite{zhao1989}Zhao {\it et al.} 1989, \markcite{leda1995}Ledlow \& Owen 1995). These results seem counterintuitive, as all current radio models require an intracluster medium (ICM) to confine the radio plasma and associated B-fields. Furthermore, cluster galaxy radio jets are frequently bent into U-shaped head-tail morphologies which are believed to result from dynamic bending caused by transonic movement through the ICM. Thus, the existence of very extended radio sources associated with galaxies apparently outside of an ICM is difficult to understand.

Analysis of {\it Einstein} data showed that the X-ray emitting gas in clusters is commonly distributed in clumps (e.g., \markcite{mohr1993}Mohr {\it et al.} 1993) within larger regions of extended emission. Radio galaxies are often correlated with the positions of these X-ray clumps (\markcite{burn1994}Burns {\it et al.} 1994), although there is some debate over the cause of this correlation (\markcite{edge1995}Edge \& R\"ottgering 1995). While the radio emission is produced by synchrotron radiation, there are several possible origins for the X-ray emission. If it originates from free-free radiation of hot ICM gas, it would be extended and could provide pressure confinement for the radio plasma. Extended X-ray emission could also arise from inverse Compton scattering of microwave background photons by the relativistic electrons generating the synchrotron emission (\markcite{daly1992}Daly 1992), though such emission would be quite weak (\markcite{burn1994}Burns {\it et al.} 1994). Another viable option for compact X-ray emission is non-thermal emission from active galactic nuclei (AGN). This X-ray emission would be unresolved and coincident with the core of radio emission. The question then becomes: are radio galaxies associated with real clumps in the ICM or do the peaks in X-ray emission arise from the galaxies themselves? Arguments for each progenitor abound in the literature (for extended emission from cluster gas, see e.g. \markcite{feig1983}Feigelson \& Berg 1983; for point emission from AGN, see e.g. \markcite{fabb1984}Fabbiano {\it et al.} 1984). Some sources may even exhibit both point-like and extended emission, possibly resulting from cooling inflows of the hot cluster gas (e.g., see \markcite{birk1993}Birkinshaw \& Worrall 1993 for a discussion of NGC6251). It seems likely that different specific circumstances require different explanations (\markcite{rhee1994}Rhee {\it et al.} 1994, \markcite{burn1994}Burns {\it et al.} 1994). 

In this paper, a complete nearby sample ($z \leq 0.06$) of 25 radio galaxies are examined in an attempt to characterize their environments. X-ray emission in the vicinity of the galaxies is investigated through use of the {\it ROSAT All-Sky Survey} (RASS) and {\it ROSAT} pointed observations. This analysis has two components: first, the determination of whether the radio galaxies are also significant X-ray sources; and second, whether this emission is extended or unresolved. Luminosities are also determined for the purpose of investigating the known correlation of cluster richness and X-ray luminosity. Optical data from the Digitized Sky Survey (DSS)
\footnote{The images on these discs are based on photographic data processed into compressed digital form at the Space Telescope Science Institute (STScI) under U.S. Government grant NAG W-2166. Copyright (c) 1993, 1994, Association of Universities for Research in Astronomy, Inc.}
 are used to determine spatial two-point correlation functions about each radio source to quantify the galaxy clustering. A list of potential poor clusters is thereby identified as those sources with significant extended X-ray emission and high optical spatial correlation coefficients. Furthermore, the specific environments of Fanaroff-Riley Class I and II (FR I and FR II, respectively; \markcite{fana1974}Fanaroff \& Riley 1974) radio sources are compared to help elucidate the nature of the origins of these two types of sources. Throughout this paper, values of $H_o=75$ km/sec/Mpc and $q_o=0.1$ will be adopted.

In Section 2, the data analysis and procedure are summarized. This summary includes the selection of the sample in addition to specific X-ray data reduction techniques. A brief overview of spatial two-point correlation functions is provided with a description of the assumptions we have used in our calculations. In Section 3 the results are summarized, and Section 4 provides a discussion of these results which focuses on correlations between radio, X-ray, and clustering properties. The results and conclusions are summarized in Section 5.

\section{Data Analysis and Procedure}
\subsection{Radio}

The sample was chosen from the 3CRR catalog (\markcite{lain1983}Laing, Riley, \& Longair 1983, hereafter LRL) and the Wall \& Peacock all-sky survey (\markcite{wall1985}Wall \& Peacock 1985, WP). The LRL sample is complete to a flux density limit $S_{178MHz} \ge 10 ~ Jy$ for sources smaller than 10' in extent. The WP survey was performed at a higher frequency than the LRL survey to avoid bias against flat-spectrum sources, and includes Southern sky sources and sources of larger extent. It is complete to $S_{2.7GHz} \ge 2 ~ Jy$ over the entire sky for which $|b| \ge 10^o$, except for regions near the Magellenic Clouds. From this combined survey, all radio sources with $z \leq 0.06$ and which were not cataloged members of Abell clusters \markcite{aco1989} (Abell, Corwin, and Olowin 1989, hereafter ACO) were selected.\footnote{ 3C66B is potentially a member of ACO 347, a richness class 0 cluster. It lies at a projected distance of about 1.5 $h_{75}^{-1}$Mpc from the center of this cluster, and its recessional velocity does not rule out its being a cluster member (z=0.022 vs. z=0.019 for the cluster). As will be shown later, 3C66B resides in its own separate region of X-ray emission and its projected distance from the core of ACO 347 will not greatly contaminate its $B_{gg}$ value.} The resulting 25 radio galaxies are listed in Table 1 along with their Fanaroff-Riley classification from \markcite{owen1989}Owen \& Laing (1989), redshift, and optical position in J2000 coordinates. Of the sample, 16 galaxies correspond to FR I sources and 6 correspond to FR II sources. The remaining three - DA240, 3C310, and NGC6251 - are transitional FR I/II sources, with the first two being classified as Fat Doubles and the third a Twin Jet. Based on the conclusions of \markcite{owen1991}Owen \& White (1991), which noted that the FR I/FR II break depends on optical as well as radio luminosity, 3C310 and NGC6251 are grouped with the FR I sources and DA240 with the FR II sources. The 1.4GHz radio powers are listed in Table 5.
\bigskip

\placetable{tbl-1}

\subsection{X-Ray}

Two-degree square fields centered on each radio source were taken from the RASS. The RASS was performed between August 1990 and February 1991, covering an effective energy range of 0.1 to 2.4 keV with a resolution of $\sim 40 \%$ at 1 keV. Exposure times vary for different locations on the sky as a result of ROSAT's elliptical orbit which provided longer exposures at the ecliptic poles, but in general are on the order of hundreds of seconds. For a more thorough description of the RASS, see \markcite{voge1993}Voges (1993). The images correspond to the energy range 0.5 - 2.0 keV. In addition, a search of the {\it ROSAT} archives found eleven of the sources (3C31, 3C66B, 3C296, 3C305, NGC6109, NGC6251, 2152-69, 3C442A, 3C449, 3C88, and Pictor A) were either the target of or were within the field of publicly-available (including some sources which lay outside the central ring of the PSPC-B rib structure) pointed PSPC-B observations. For both RASS and pointed images, photometry was performed to determine X-ray luminosities and characterize the emission as extended or unresolved.

Photometry of the RASS images was performed with the task IRING found within the NRAO's Astronomical Image Processing System (AIPS). IRING sums counts within annuli about the given central position of the source, which was determined by a best-fit Gaussian to the smoothed X-ray image. The chosen aperture corresponded to a linear radius of 250 $h_{75}^{-1}$kpc, and sixteen equally-spaced annuli were used within this aperture. The background was estimated using border pixels around the image, rejecting values greater than $3 \sigma$ above the mean. In converting net count rates to luminosities, energy conversion factors (ECFs) which assume a 1 keV Raymond-Smith plasma (\markcite{raym1977}Raymond \& Smith 1977) of 0.3 solar metallicity were used. Corrections for photoelectric absorption based on hydrogen column densities from \markcite{star1992}Stark {\it et al.} (1992) were also applied.

Photometry of available pointed observations follows the same procedure, with two notable differences. Due to the position-dependent sensitivity of the PSPC-B, background counts were determined using three equivalent-sized apertures located in similar regions of the detector. Care was taken to avoid including rib structures in these background estimates. The ECFs include the same assumptions as those used for RASS data, with the exception of the different response function for the PSPC-B detector. In all cases, the luminosities calculated from these pointed observations were consistent with those derived from the RASS.

To differentiate between point and extended sources, a point source profile was created for the RASS images using point sources located within the fields of the 25 sources. This profile is depicted in Figure 1, and is well-described by a Gaussian, but with a long tail. Comparison of the profile of each source with the generated point response function was performed to differentiate between point and extended sources. Two X-ray sources lacked adequate S/N for characterization (NGC1692 and 3C88), although the latter of these was also present in pointed observations so the differentiation was achieved. For pointed observations, the task PCRPSF within the FTOOLS software package was used to generate theoretical point response curves for comparison with radial profiles. This task accounts for the variation in the point response function of the PSPC-B with off-axis angle and energy.

\placefigure{fig1}

\subsection{Optical}

Optical data from the Digitized Sky Survey (DSS) were used to determine a quantitative measure of the degree of clustering. Abell characterized clusters by simply counting the number of galaxies within a magnitude range $m_3$ to $m_3+2$, where $m_3$ is the apparent magnitude of the third-brightest galaxy. These counts were performed after correcting for the background and corresponded to galaxies within a radius defined by a set angular distance divided by the assumed redshift of the cluster (a linear radius of about $2.0h_{75}^{-1} Mpc$ for nearby clusters). Using this procedure he established richness classes of clusters based on their number of members. A more statistical measure of clustering is provided by the spatial two-point correlation function, which measures the probability in excess of random of finding two objects in a given volume of space. It is frequently characterized as a coefficient multiplied by a power-law in the separation. This power-law holds over many scales, meaning the coefficient represents a measure of the degree of clustering (and is frequently referred to as the amplitude of the correlation function). \markcite{ande1994}Andersen \& Owen (1994, AO94) and \markcite{yeel1999}Yee \& Lopez-Cruz (1999, YLC99) demonstrated that there is a correlation between this coefficient and Abell's richness class.  Hence, we rely on the spatial two-point correlation coefficient as our means to assess clustering. A brief background is provided (for a more thorough review, see \markcite{pjep1980}Peebles 1980), followed by a more specific procedure which parallels that presented by \markcite{ande1994} AO94.

\subsubsection{The Two-Point Correlation Coefficient, $B_{gg}$}

Given the presence of one object, the two-point correlation function, $\xi (r)$, is defined in terms of the probability of finding another object within a volume element $\delta V$:
\begin{equation}
\delta P = n \delta V (1 + \xi (r))
\end{equation}
where $n$ is the mean volume density of objects and the volume is described by a radius vector, $r$. Integrating this expression simply gives the number of galaxies counted within the volume. Note that for $\xi (r) = 0$ the distribution is random and the expected number of objects is simply the mean density times the volume. Positive values of $\xi (r)$ therefore denote clustering in excess of random.

\markcite{grot1977}Groth and Peebles (1977) determined that the distribution of galaxies was well-fit by a power law such that
\begin{equation}
\xi (r) = B_{gg} r^{-\gamma}
\end{equation}
where $\gamma \sim 1.77$. $B_{gg}$ is therefore called the spatial two-point correlation coefficient and represents a measure of the degree of clustering of a system. Groth and Peebles found that, on average, galaxies exhibit clustering on scales up to $\sim 12~h_{75}^{-1}Mpc$ with $B_{gg}\approx26$ $h_{75}^{-1.77}Mpc^{1.77}$.

Lacking three-dimensional data on the distribution of galaxies, $B_{gg}$ can be estimated from the angular distribution. Analagous to the spatial two-point correlation function is the angular two-point correlation function, $\omega (\theta)$ defined such that $\omega (\theta) = A_{gg} \theta ^{1 - \gamma}.$  Substituting this into the angular equivalent of Equation 1 and integrating yields:
\begin{equation}
A_{gg} = {1.23 \over 2 \pi {\cal N} \theta ^{1.23}}(\bar N - \pi \theta ^2 {\cal N})
\end{equation}
where ${\cal N}$ is the mean surface density of galaxies and $\bar N$ is the counted number of galaxies brighter than some limiting magnitude, $m_o$. Assuming little evolution of the correlation function and the galaxy luminosity function, the angular correlation coefficient is related to the spatial one via:
\begin{equation}
B_{gg} = {A_{gg}{\cal N} \over I_{\gamma} \phi (m_o,z)}\left({D_L \over (1+z)^2}\right)^{\gamma - 3}
\end{equation}
with $D_L$ representing the luminosity distance; $\phi (m_o,z)$ the galaxy luminosity function (LF); and $I_{\gamma}$ is a constant resulting from the integration (equal to 3.78 for $\gamma = 1.77$).

\subsubsection{Procedure for Estimating $B_{gg}$}

The above discussion shows that there are several parameters which must be specified in order to determine values of $B_{gg}$ for the radio galaxies in our sample. The primary parameters include a LF, a limiting magnitude, and an angular counting radius. Further issues include the choice of a value for $\gamma$ and determination of the mean surface density of galaxies, ${\cal N}$. We have relied upon signal-to-noise (S/N) criteria in our determination of these parameters and hence $B_{gg}$. An Abell LF (\markcite{abel1962} Abell 1962; the LF is approximated as two power laws meeting at $M^*$) is assumed and we count down to the corresponding value of $m^*$, corrected for extinction. Our angular counting radius varies slightly, but is typically around $800h_{75}^{-1}kpc$. A summary of our chosen parameters is included in Table 2, and a detailed description of the procedure may be found in \markcite{ande1994} AO94. Galaxies were selected by eye from the DSS and we performed our own photometry. To eliminate mis-classifications of stars as galaxies we performed Gaussian fits to the objects and rejected objects with stellar profiles.

\placetable{tbl-2}

\subsection{Comparison of $B_{gg}$ Values with Other Studies}

\subsubsection{Results for Abell Clusters}
Several other studies have used $B_{gg}$ to quantify clustering. Of these, \markcite{pres1988} Prestage \& Peacock (1988; PP88) also investigated a few of the radio sources in this sample. More recently, \markcite{yeel1999} YLC99 calculated $B_{gg}$ values for Abell clusters and argued for their use as a more robust measure of cluster richness. Table 3 presents the derived $B_{gg}$ values by richness class for each of these two studies. For comparison, the results of \markcite{ande1994} AO94, who used essentially the same procedure as we have in this paper, are also presented. As there are clear discrepancies across these studies, some discussion is warranted.

\placetable{tbl-3}
 
The three studies may be normalized by applying scaling relations based on their chosen values of $H_o$ and LF. For example, as $B_{gg}$ $\propto$ $r^{-1.77}$, $B_{gg}$ values computed with $H_o$ = 50 km/sec/Mpc (PP88, YLC99) will be about a factor of two greater than those computed with $H_o$ = 75 km/sec/Mpc (AO94). The remaining large-scale differences can be explained by noting the differences in assumed LFs. Since $B_{gg}$ is inversely proportional to the LF, its shape and normalization have dramatic effects on the resulting values. The size of this effect depends on the difference in the LFs at the limiting magnitude used for galaxy counting. For example, the LF and associated parameters used in \markcite{yeel1999} YLC99 result in higher values of $B_{gg}$ by an additional factor of two over those in \markcite{ande1994} AO94. Applying this net factor of four to the AO94 values brings them into line with those of \markcite{yeel1999} YLC99.

As alluded to in Section 2, a number of additional parameters lead to smaller discrepancies in calculated $B_{gg}$ values. A discussion of these and their relative effects may be found in \markcite{yeel1999} YLC99.

\subsubsection{Application to This Study}
As the radio galaxies analyzed in this study are not members of rich clusters, the optimal choices of limiting magnitude and counting radius may not be the same as those for evaluating $B_{gg}$ for Abell clusters. The basic arguments are that by counting to too faint a magnitude or too large an angular extent, the background counts will begin to dominate over the cluster counts and thereby reduce the S/N of the result. To investigate these parameters we created a computer model. The `clusters' were represented by King profiles (\markcite{king1962} King 1962; $\beta$ models) of galaxy density, with luminosities obeying the Schechter LF \markcite{sche1976} (Schechter 1976) used in \markcite{yeel1999} YLC99. Models with different core radii and richnesses were run to explore a range of cluster redshifts and background counts. The rich clusters were modeled by a King function with core radius equal to $100h_{75}^{-1}kpc$ and $\beta$ equal to one, values suggested in \markcite{adam1998}Adami {\it et al.} (1998). The environments around the radio galaxies were modeled with a smaller core radius of $50h_{75}^{-1}kpc$, and the richness was varied to find a model which approximately matched the representative $B_{gg}$ values derived for the actual radio galaxies.

Evaluation of the best choice of limiting magnitude and counting radius was made by the combination which produced the highest S/N for that model. With core radius and richness fixed, isoclines of constant S/N for varying limiting magnitude and counting radius result. Figure 2 depicts the results for a both a model radio galaxy environment and a model ACO cluster of richness class 1. Contours in the figure depict isoclines of constant S/N from 50$\%$ to 100$\%$ of the peak S/N, in increments of 10$\%$. The model results clearly show that for richer environments, the S/N improves by counting out to greater radii and fainter limiting magnitudes. The figure also shows that the S/N rises fairly rapidly and reaches a broad plateau. This means that a reasonably wide range of counting radii and limiting magnitudes produce S/N which are better than 90$\%$ of the peak value, and confirms the results of \markcite{yeel1999}YLC99 which demonstrate $B_{gg}$ is reasonably insensitive to these parameters. The optimal values of counting radius and limiting magnitude are presented in Table 4 for average richness class 0, 1, and 2 clusters and the average radio galaxies of this study.

\placefigure{fig2}

\placetable{tbl-4}

These models confirm that the procedure of \markcite{ande1994}AO94 applied in this study is a good choice for the poorer environments of the sample radio galaxies. Compared to Abell clusters, higher S/N for the radio galaxies' $B_{gg}$ values is obtained by counting to brighter magnitudes and smaller counting radii. While the \markcite{ande1994}AO94 parameters are still not quite optimal for the radio galaxy environments in this sample, they usually result in only a small reduction in S/N of 10$\%$ or less. The \markcite{ande1994}AO94 parameters lead to greater reduction in S/N as one moves toward richer environments. For the richest radio galaxy environment in our sample (NGC1692), the reduction amounts to about 30$\%$ and for an Abell cluster of richness class 2 it is on the order of 50$\%$. For comparison, the parameters adopted by \markcite{yeel1999}YLC99 would reduce the S/N in the $B_{gg}$ for a typical radio galaxy in this sample by around 30$\%$. Consequently, we argue that the use of the \markcite{ande1994}AO94 parameters is the better choice for this study.

The parameters of \markcite{yeel1999}YLC99 are, however, a good choice for the richer environments of Abell clusters. While our model results suggest that \markcite{yeel1999}YLC99 would improve the S/N in their estimates by using a slightly brighter limiting magnitude and larger counting radius, the net improvement is less than 10$\%$. Furthermore, our models confirm that for the fixed counting radius of about 700 $h_{75}^{-1}$kpc used by \markcite{yeel1999}YLC99, they achieve the best results by counting down to their value of $m^*$+2.

\section{Results}

Most of the radio galaxies were associated with significant detections in the X-ray. Of the 18 FR I sources, only 3 were not detected at the $\ge 3 \sigma$ confidence level (3C76.1, 3C293, and 3C386). Five of the 7 FR II sources were significant detections, with only DA240 and 3C98 not being detected. Table 5 lists the X-ray luminosities for each field, derived using a 250 $h_{75}^{-1}$kpc radius aperature. The X-ray luminosities for all of the detected sources lie at the high end or above those expected for normal E/S0 galaxies (\markcite{fabb1989}Fabbiano 1989). The reported X-ray luminosities and associated S/N for several of the detected sources appear low in Table 5 as a result of their being fairly weak, compact sources whose signal is overwhelmed by the background in the 250 $h_{75}^{-1}$kpc aperature.

The X-ray extents of the sources are clearly essential to understand their environment. Consequently, radial profiles for the significant X-ray detections were compared with the point response function determined for the RASS. Figure 3 illustrates RASS profiles for 3C31, 3C66B, NGC6251, and Pictor A. The first two are clearly extended sources while Pictor A is unresolved. NGC6251 looks like a point source at small radii, but shows evidence for weak, extended emission at larger radii. This is consistent with the findings of \markcite{birk1993}Birkinshaw \& Worrall (1993), who suggest NGC6251 is best represented by a point source with a faint X-ray halo. In all, eleven of the detected sources were characterized as extended with three others being unresolved (3C78, 2152-69, and Pictor A). The remainder of the detected sources lacked adequate signal for characterization (3C278, 3C296, 3C305; and 0131-36, 3C88, 1637-77). These results were confirmed with profiles derived from pointed observations, which also showed 3C88 to be unresolved.

\placefigure{fig3}

A simple estimation of the radial extents was performed by noting the radius at which the source signal was less than $3 \sigma$ above the background. To eliminate the possibility of chance fluctuations leading to incorrect characterization of source size, this criterion had to be met in three successive annuli. If this occurred, the radial extent was assigned as the linear distance corresponding to the last annulus significantly above the background. As it is inaccurate to assume that the X-ray distributions are symmetric and centered on the radio source, and our linear resolution is dependent on the redshift of the source, these extents should be considered as rough approximations. To illustrate this and gain a better understanding of the nature of the X-ray emission, overlays of X-ray contours onto DSS optical images are provided in Figure 4. In addition to the eleven extended sources, an overlay for 2152-69 (an unresolved source with high $B_{gg}$) is included. Figure 4b shows that 3C66B is associated with its own region of extended X-ray emission, with the X-ray emission associated with the core of ACO 347 lying well to the south of the image. For this reason, we justify its inclusion in our sample of non-cluster radio galaxies. 3C353 was removed from the list of extended sources as it is not within an extended X-ray source but lies next to one (see Figure 4k; also \markcite{sieb1996}Siebert {\it et al.} 1996). Consequently, averaging over the fluxes contained in successive annuli gives the false appearance of extended X-ray emission about the source. 

\placefigure{fig4} 

Overlays of X-ray contours onto radio images for five of the larger radio sources (3C31, 3C66B, 3C310, 3C442A, and 3C449) are depicted in Figure 5. In each of these sources, the extent of the radio emission is approximately the same as the extent of the X-ray emission.

\placefigure{fig5}

\placetable{tbl-5}

In optical measures of clustering, 11 of 18 ($61 \%$) FR I sources exhibit $B_{gg}$ values $\ge 1\sigma$ in excess of that expected for a typical galaxy ($B_{gg}\approx26~h_{75}^{-1.77}Mpc^{1.77}$; Groth \& Peebles 1977, determined as the average for all galaxies present in the Lick survey). Only 2 of the 7 ($29 \%$) FR II sources were significant at this level. One of these two sources was 3C353, which as noted previously in the X-ray discussion is the result of a superposition of the radio galaxy with an offset X-ray and optical group. Results for three sources (IC2082, IC4296, and 2104-25) located within ACO clusters are also included in Table 5. These sources were included initially in the sample to provide a consistency check on the results.

The evidence for clustering is even stronger than suggested by the $B_{gg}$ values. For poor clusters such as those of this study, low values of $B_{gg}$ do not necessarily imply a lack of physical association for galaxies in the vicinity of the radio galaxy. This is because the $B_{gg}$ calculation for a close collection of galaxies may be dominated by the background. The presence of clustering was also investigated with Poisson statistics. Given our assumption of a uniform distribution of field galaxies $\propto 10^{0.6m}$, we can easily determine the expected number of field galaxies within each radio galaxy's survey area. This may be compared to the actual number of galaxies counted to test the null hypothesis that there is no cluster present and the galaxy counts represent only chance fluctuations in the field galaxy distribution. Performing this simple test, 18 of the 25 galaxies represented clustering in excess of the field galaxy distribution at the $95\%$ significance level or greater. This number breaks down into 15 of 18 FR I galaxies and only 3 of 7 FR II galaxies (including 3C353). The counted number of members and expected background counts per field are also listed in Table 5.

\section{Discussion}

\subsection{Radio Galaxies as Cluster Members}
 
The primary result of this investigation is that many of these nearby radio galaxies do, in fact, reside in cluster environments. As one would expect based on their not being cataloged members of clusters, these cluster environments are poorer than those prescribed by Abell. The data backing this claim are the high incidence of significant X-ray emission associated with the radio galaxies and their significant optical clustering above that expected for the field. To demonstrate this point, Table 3 includes average values of X-ray luminosity and $B_{gg}$ for Abell clusters of richness class 0, 1, and 2. These data are also included in Figure 8, where it can be seen that the radio galaxies form a natural extension of the classical Abell clusters in regards to X-ray luminosity and optical clustering.

The difference between the FR I and FR II sources is striking. All ten of the extended X-ray sources are FR I sources, whereas none of the seven FR II sources are resolved. Furthermore, most of the FR I sources represent significant associations of galaxies above the background (15 of 18), including all ten of the extended X-ray sources. Conversely, only three of the seven FR II sources are indicative of significant clustering above an expected field distribution. The significance of such clustering merely strengthens this argument, as only two FR II sources possess $B_{gg}$ values in excess of that expected for a typical galaxy. This is even including 3C353, which can be seen to lie at the edge of an optical group of galaxies coincident with X-ray emission.

Consequently, it is more correct to say that the FR I radio galaxies represent a low richness extension of the Abell clusters. They are frequently associated with extended X-ray emission suggestive of an ICM, and possess a significant excess of nearby galaxies above that expected for the field. The strong radio-emitting FR II galaxies appear to reside in environments more consistent with the field. They are either non-detections in the X-ray or possess unresolved X-ray emission. This suggests a more internal mechanism for their X-ray emission, namely AGN activity. Thus, FR I's are more likely to be associated with some amount of clustering than FR II's. These results confirm those of several prior studies, including \markcite{long1979}Longair \& Seldner's (1979) pioneering work which noted that FR I's were more likely to be associated with regions of greater clustering. \markcite{pres1988} PP88 strengthened this result and added that FR II sources avoid high richness environments at low redshift whereas at higher redshifts FR I sources and FR II sources are found in similar environments.

It is also interesting to note the overlap in the X-ray and optical results: of the five X-ray non-detections, only one (DA240) represents a marginally significant excess of galaxies above the field. The overlap is exact for the FR I radio galaxies: of the 18 FR I sources, fifteen possess significant X-ray emission and an excess of galaxies above the field, whereas the remaining three possess neither (3C76.1, 3C293, and 3C386).

\subsection{Correlations and Their Implications for Environment}

Prior studies have noted that there appears to be no correlation between radio power and richness for Abell clusters (e.g., \markcite{zhao1989}Zhao {\it et al.} 1989, \markcite{leda1995}Ledlow \& Owen 1995). Correlation analyses between radio power at 1.4 GHz ($P_{1.4GHz}$) and optical clustering ($B_{gg}$) were performed to test this result for our sample. Kendall's $\tau$ for this correlation was determined for the entire sample as well as subsets defined by radio morphology (see Table 6 and Figure 6). In performing these tests, censoring of data was used when $B_{gg}$ values were considered to be upper limits. These upper limits corresponded to the $B_{gg}$ value at which any underlying galaxy population would be significantly in excess of the expected field counts at the $95\%$ confidence level. For reference, these upper limits are all roughly equal to the \markcite{grot1977} Groth \& Peebles (1977) average galaxy $B_{gg}$ value of $\approx26~h_{75}^{-1.77}Mpc^{1.77}$. We cannot make any statements about the correlation between radio power and richness for our sample as a whole. However, breaking the sample into different classes provided interesting insight, though caution must be exercised in evaluating the correlations as the sample sizes are small. Performing the correlation analysis for only the extended FR I sources indicated a potential correlation between $P_{1.4GHz}$ and $B_{gg}$, at a probability of $98.4\%$. This result suggests that the presence of an ICM as indicated by extended X-ray emission does have a significant effect on FR I radio sources, in accordance with present theories of such sources.

\placetable{tbl-6}
\placefigure{fig6}

The importance of X-ray gas was further investigated by analyzing potential correlations between $P_{1.4GHz}$ and $L_x$. The results of the Kendall's $\tau$ tests (again with censoring) are presented in Table 6, and Figure 7 graphically depicts the data. As expected from the $P_{1.4GHz}-B_{gg}$ correlations, little can be said about potential correlations for the entire sample or the FR I sources in particular. Correlations between $P_{1.4GHz}$ and $L_x$ appear to exist for both the FR II and extended FR I sources (at $96.4\%$ and $98.8\%$, respectively), though this is most likely the result of selection bias. As our sample is derived from flux-limited all-sky surveys, the radio powers and redshifts are very strongly correlated. Since $L_x$ is also redshift dependent, it is not surprising that it and $P_{1.4GHz}$ appear correlated. Given this caveat, Figure 7 shows different trends for the FR II and extended FR I sources, which is a further suggestion of different emission mechanisms. A qualitative understanding of the relationship between the radio and X-ray for the FR I sources may be derived from Figure 5. In this figure, radio images of five of the extended FR I sources are shown with X-ray contour overlays. In each case, the extents of the radio and X-ray emission are consistent with one another. 

\placefigure{fig7}

Correlation between $L_x$ and $B_{gg}$ is less straight forward. For the entire sample, the two variables are likely to be correlated (greater than a $98\%$ chance that a correlation is present; see Table 6). This can be interpreted as richer systems being more likely to have associated X-ray gas. Broken down by radio morphology, this correlation appears to hold for the entire FR I portion of the sample (again a greater than $98\%$ probability) while there is no significant correlation for the FR II sources. While one would expect $L_x$ and $B_{gg}$ to be strongly correlated for the extended FR I sources, the results of the Kendall's $\tau$ test do not support this. Figure 8 plots $B_{gg}$ vs. $L_x$ for each type of source.

\placefigure{fig8}

\section{Conclusions}

A thorough study of the environments of a statistical sample of nearby radio galaxies has been presented. By focusing on both X-ray emission and optical clustering, stronger conclusions may be reached than for studies which address only one of these wavelengths. Several important conclusions are summarized as follows:

\begin{itemize}
\item Radio galaxies not cataloged to be in rich clusters are frequently associated with significant X-ray emission (20 of 25, or $80\%$ of the total sample) and galaxy clustering in excess of that expected for the field (18 of 25, or $72\%$). This suggests that powerful radio galaxies are associated with cluster environments, even if these clusters are too poor to be classified as such in well-known catalogs.

\item Of the radio galaxies with associated X-ray emission, many (10 of 20) are resolved by {\it ROSAT}. This suggests the presence of an ICM around these sources, which is consistent with current models of extended radio sources.

\item Though the small sample size limits strong statistical conclusions, FR I sources are more likely to lie in cluster environments. Ten of 18 ($56 \%$) of the FR I sources exhibit extended X-ray emission vs. none of the FR II sources. This same trend is reflected in the $B_{gg}$ values, with 11 of the 18 ($61\%$) FR I sources exhibiting $B_{gg}$'s significantly in excess of that expected for a typical galaxy. All of the FR I sources with significant $B_{gg}$ are also extended X-ray sources. 

\item FR II sources are more likely associated with compact X-ray sources, and hence are consistent with AGN. Five of the seven FR II sources are significant X-ray detections yet none are resolved. These results, coupled with the optical results showing FR II galaxies to generally lie in field-like environments, suggest FR II galaxies are governed more by influences within the galaxies and less by the presence of a cluster environment. 

\item While FR II sources may lie in regions of modest clustering (2 of 7 have significant values of $B_{gg}$), they do not appear to be centrally located in such clusters. This is evidenced by the fact that none of the FR II sources lie within regions of extended X-ray emission, and at least one example (3C353) can be seen to lie on the outskirts of such a region.

\end{itemize}

Hence, we have demonstrated that for this nearby sample FR I radio galaxies are associated with cluster environments, even if these environments are too poor to be classified as clusters by classical measures. Moreover, the environments of these nearby FR I and FR II sources differ, with FR I sources more likely to be associated with X-ray gas and regions of higher optical clustering. Thus, we can conclude that a more significant ICM is present for FR I radio sources. 

Other studies aimed at cluster radio galaxies have found that higher redshift FR I and FR II galaxies reside in similar environments, and suggested that this similarity appears to break down at lower redshifts. This study has strengthened this latter claim, thereby adding importance to the issue of evolution of FR I and FR II environments. Why are FR I and FR II galaxies found in similar cluster environments at high redshift while the FR II galaxies appear to avoid cluster environments at low redshift? That is, what evolutionary factors cause this change in the environment of FR II radio galaxies? Such questions need to be the topic of future studies.

Even within this study's sample of nearby FR I galaxies there are additional unanswered questions. If a cluster environment seems to be so important for FR I galaxies, what is the explanation for the three FR I sources of the sample without X-ray emission or significant clustering? Are their radio properties and morphologies somehow different than the other FR I galaxies of the study?

Answering these questions requires more evidence of the conclusions reported in this paper. The measures of clustering we have used are statistical in nature. More direct evidence of clustering can be obtained through collection of spectra of the galaxies in these environments. From this information, velocity dispersions and subsequently mass estimates can be made. We have undertaken such a project, and thereby hope to better determine whether these systems do in fact represent poor clusters of galaxies. This information will demonstrate more conclusively if the clustering environments of nearby FR I and FR II galaxies differ, and should further elucidate the differences in the environments of the sample FR I galaxies for which no evidence of clustering was found. In addition to the radio galaxies of this study, we hope to increase the size of our sample of nearby radio galaxies in order to strengthen our statistical conclusions.

\acknowledgments

The authors would like to thank MPE and its director, J. Tr\"umper, for making the RASS data available for this study. Clay Smith provided invaluable assistance by writing routines to assist in the reduction of the X-ray data. We also thank Victor Andersen, Mark Bliton, Percy G\'omez, Chris Loken, Glenn Morrison, and Elizabeth Rizza for conversations regarding this research, and Heinz Andernach and an anonymous referee for valuable comments on the manuscript. We acknowledge NASA grant NAGW-3152 to F.N.O. and J.O.B. and NSF grants AST-9317596 and AST-98 to J.O.B.
N.A.M. would like to thank the NRAO pre-doctoral program for support of this research.

\clearpage

\begin{figure}
\figurenum{1}
\plotone{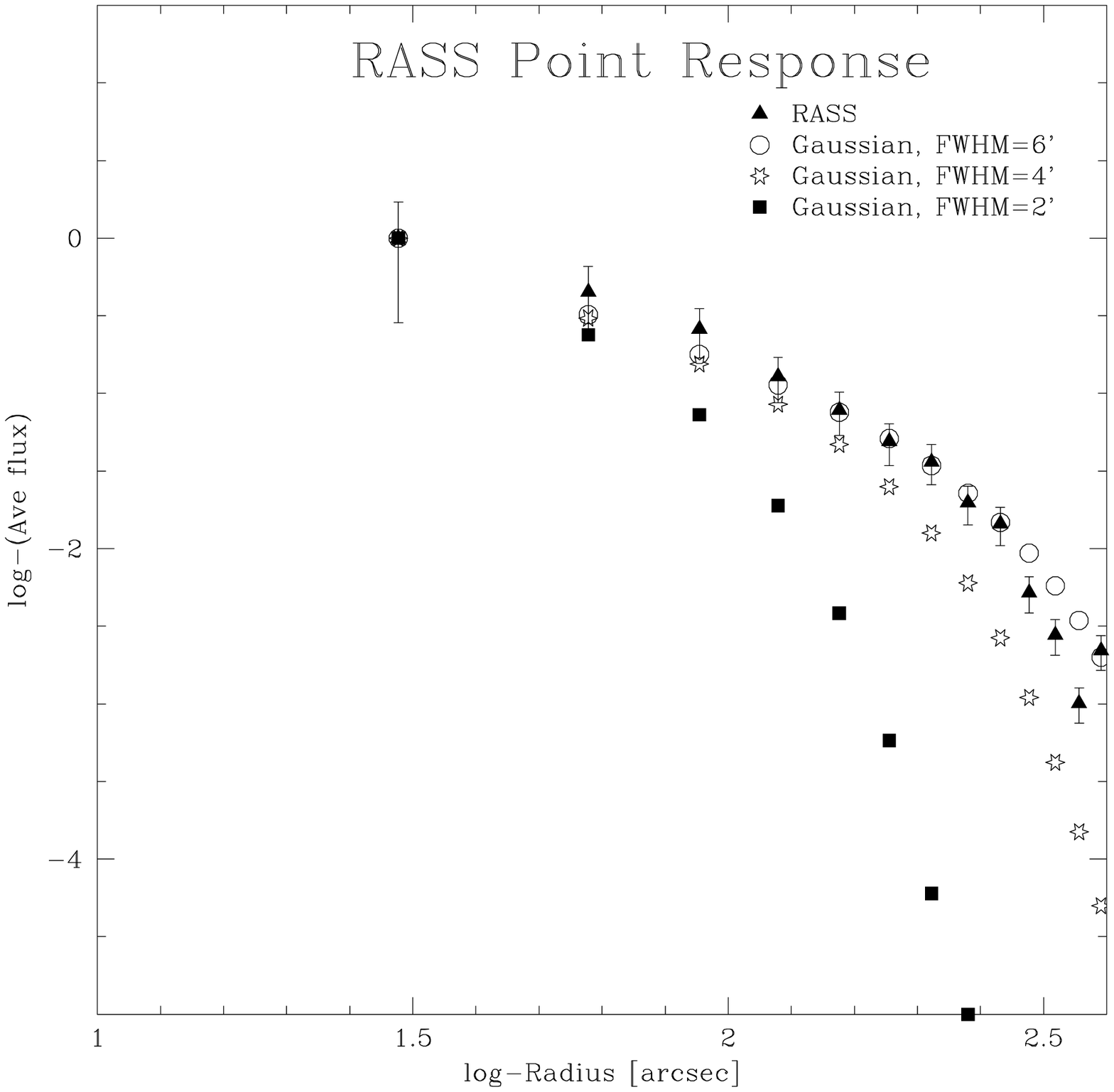}
\caption{The point response function determined for the RASS, plotted as filled triangles with $2 \sigma$ error bars. Fluxes are determined within concentric annuli which are each 2 pixels (30") in extent. Thus, the vertical axis depicts the average flux per pixel within the annulus whose outer radii is plotted on the horizontal axis. For comparison, Gaussian profiles are also plotted. These are determined using the same procedure which effectively smoothes the profile, meaning that the true FWHM of the RASS is significantly smaller than that of the best fitting curve. The FWHM for the three Gaussian curves (squares, stars, and circles) are 2', 4', and 6', respectively. As these are fits to the data post-smoothing, the actual FWHM of the point response function is significantly smaller.}
\end{figure}

\begin{figure}
\figurenum{2}
\plotone{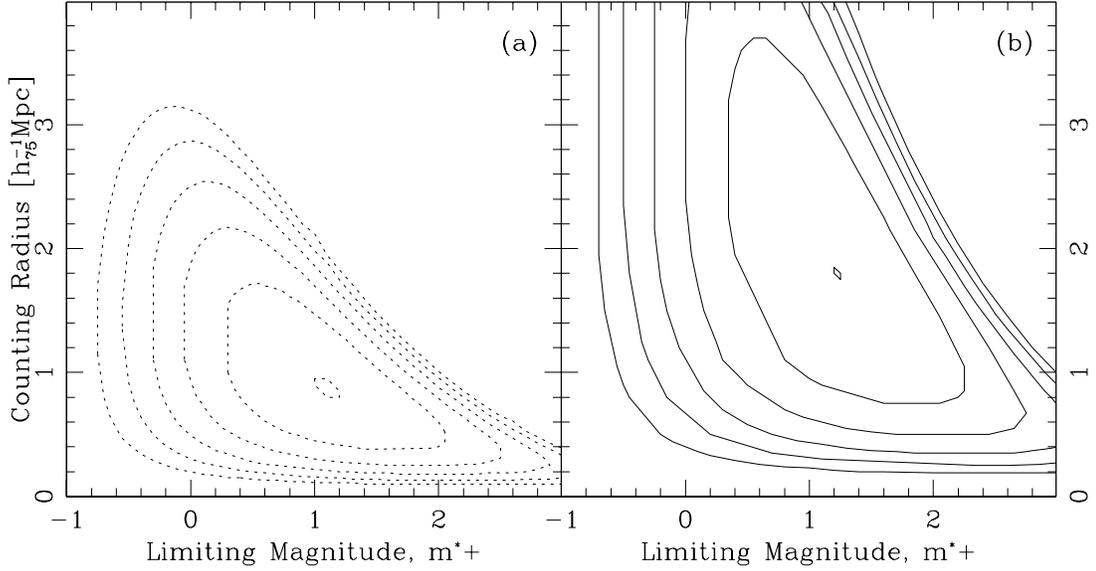}
\caption{Counting radius and limiting magnitude to achieve optimal S/N for (a) a model radio galaxy environment typical of this study, and (b) a model ACO richness class 1 cluster. The contours begin at a base level of $50\%$ of the peak S/N and are incremented by $10\%$. Note the brighter limiting magnitude and smaller counting radius that produce optimal results for the poorer environments around the sample's radio galaxies.}
\end{figure}

\begin{figure}
\figurenum{3}
\plotone{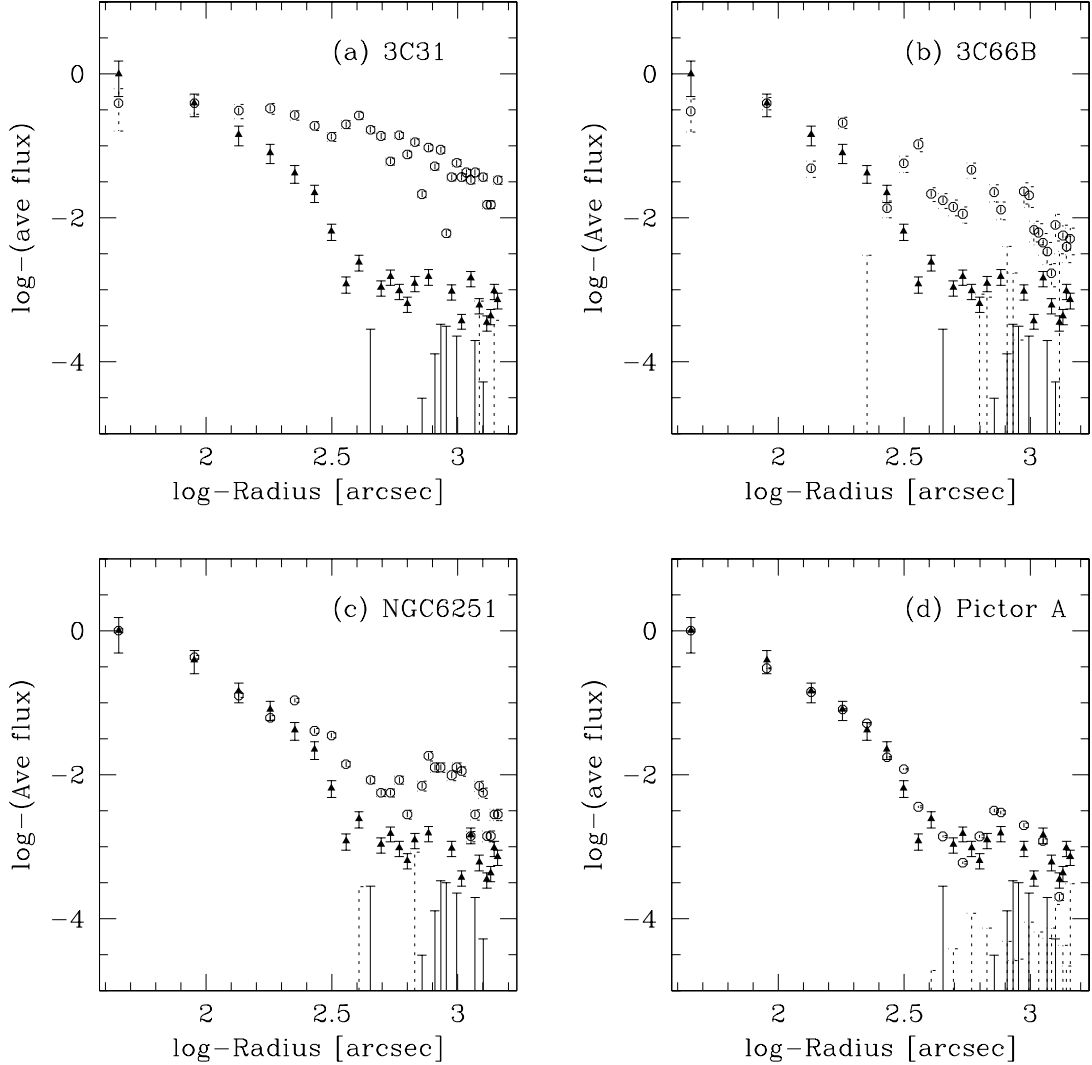}
\caption{Normalized radial profiles for (a) 3C31, (b) 3C66B, (c) NGC6251, and (d) Pictor A. Open circles represent the source profiles, and filled triangles represent the point response of the detector. The error bars represent $2 \sigma$, with dashed lines for the sources and solid lines for the point response. The error bars off the bottom of the plot demonstrate the radius at which noise dominates. 3C31 is clearly a broad, extended source with significant emission beyond that expected for a point source. 3C66B is also extended, whereas Pictor A has a profile that closely matches the RASS point profile. Note that NGC6251 appears to be a point source at small radii, but exhibits extended emission at larger annuli.}
\end{figure}

\begin{figure}
\figurenum{4}
\caption{X-ray contours overlayed on DSS optical images. The X-ray images have been smoothed with a 2' Gaussian and background subtracted. In all cases, the contours correspond to (-3, 3, 6, 9, 12, 15, 18, 21, 24, 27, 30) times the stated background level and the greyscale has been adjusted to bring out detail. The optical positions of the radio galaxies are marked by a cross. (a) 3C31 - X-ray image taken from ROSAT pointed observation, with the base contour level being $1.2~\times~10^{-5}$ counts/sec; (b) 3C66B - X-ray image from RASS data, with contour levels being $3.1~\times~10^{-1}$ counts; (c) NGC1692 - X-ray image from RASS data, with the contour level being $1.4~\times~10^{-5}$ counts; (d) NGC5090 - X-ray image from RASS data, with the contour level being $2.3~\times~10^{-5}$ counts, (e) 3C310 - X-ray image from RASS data, with the contour level being $1.3~\times~10^{-5}$ counts, (f) NGC6109 - the X-ray image is taken from ROSAT pointed observations, with the contour level corresponding to $8.4~\times~10^{-6}$ counts/sec; (g) NGC6251 - X-ray image from ROSAT pointed observations, with the contour level corresponding to $4.8~\times~10^{-5}$ counts/sec. While the overlay suggests this is an unresolved X-ray source, Figure 4c demonstrates the presence of weak extended emission away from a strong unresolved core; (h) 3C442A - X-ray image taken from RASS data, with the contour level being $1.7~\times~10^{-1}$ counts; (i) 3C449 - X-ray image taken from ROSAT pointed observation, with the contour level being $6.1~\times~10^{-6}$ counts/sec; (j) NGC7385 - X-ray image taken from the RASS, with the contour level corresponding to $1.7~\times~10^{-5}$ counts; (k) 3C353 - X-ray image taken from the RASS, with the contour level being $2.6~\times~10^{-5}$ counts. Note that the contours for 3C353 indicate that its X-ray luminosity arises from a nearby source.}
\end{figure}

\begin{figure}
\figurenum{5}
\caption{X-ray contours overlayed on radio maps. The X-ray contours are the same as used in Figure 4. The sources are (a) 3C31 at 600 MHz. The radio image was taken with the Westerbork Synthesis Radio Telescope (WSRT) and appears in the DRAGN atlas (Leahy, Bridle, \& Strom); (b) 3C66B at 1.4 GHz. This image was taken with the Very Large Array (VLA) by Hardcastle et al. (1996); (c) 3C310 at 1.4GHz. The radio image was taken with the VLA by van Breugel \& Fomalont (1984); (d) 3C442A at 1.4 GHz. The radio image was taken with the VLA by Comins \& Owen (1991); and (e) 3C449 at 600 MHz. This image was taken with the WSRT and is available through the DRAGN atlas.}
\end{figure}

\begin{figure}
\figurenum{6}
\plotone{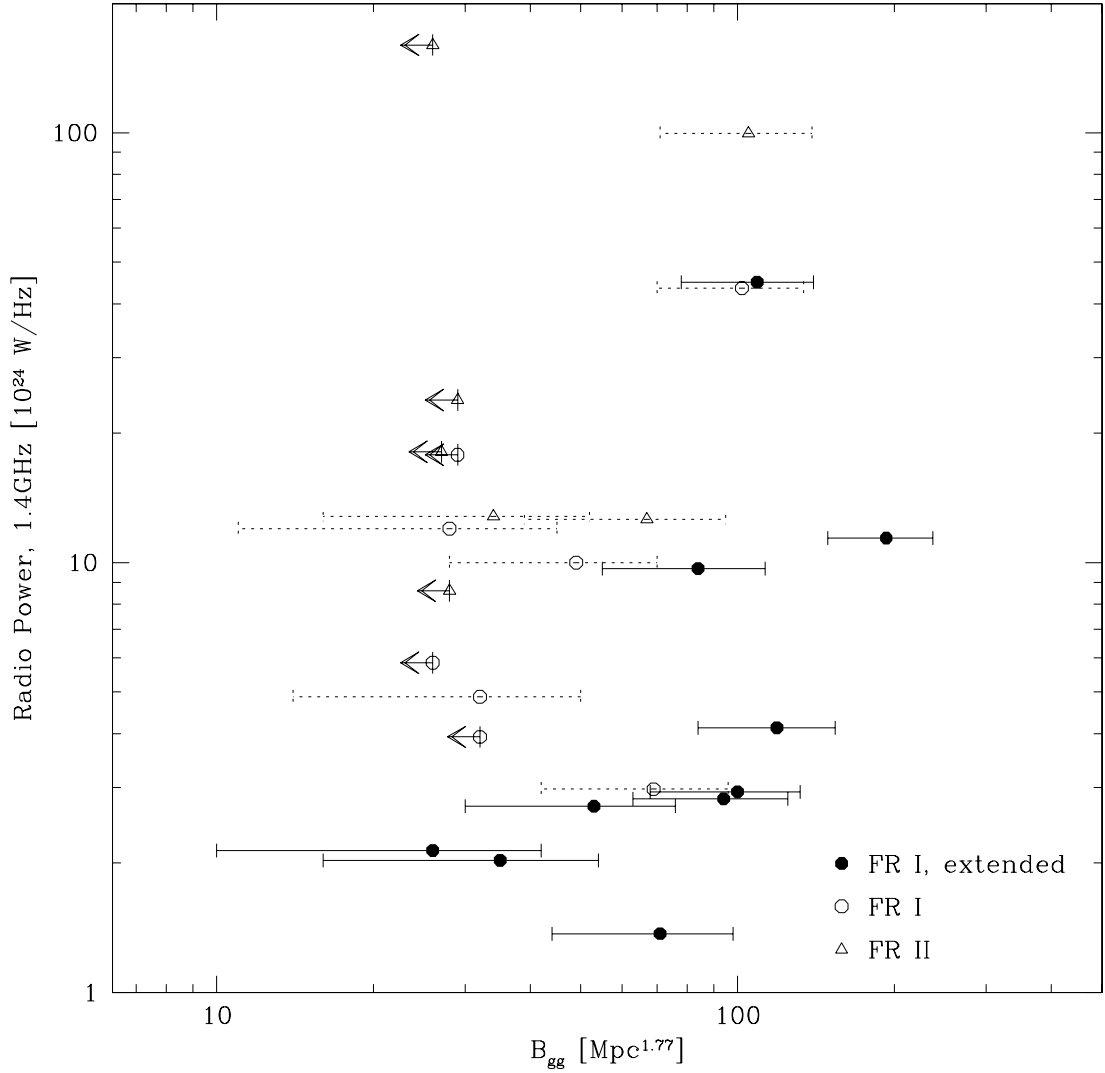}
\caption{$B_{gg}$ vs. radio power at 1.4 GHz, where $B_{gg}$ is calculated in this study and Table 1 lists the references for the radio powers. Filled circles represent FR I sources with extended X-ray emission whereas open circles are FR I sources which are unresolved by RASS. The triangles represent the FR II sources. The arrows denote upper limits, as determined by the $B_{gg}$ value at which the supposed cluster would be significantly above the background with $95\%$ confidence. Error bars for $B_{gg}$ are $1\sigma$.}
\end{figure}

\begin{figure}
\figurenum{7}
\plotone{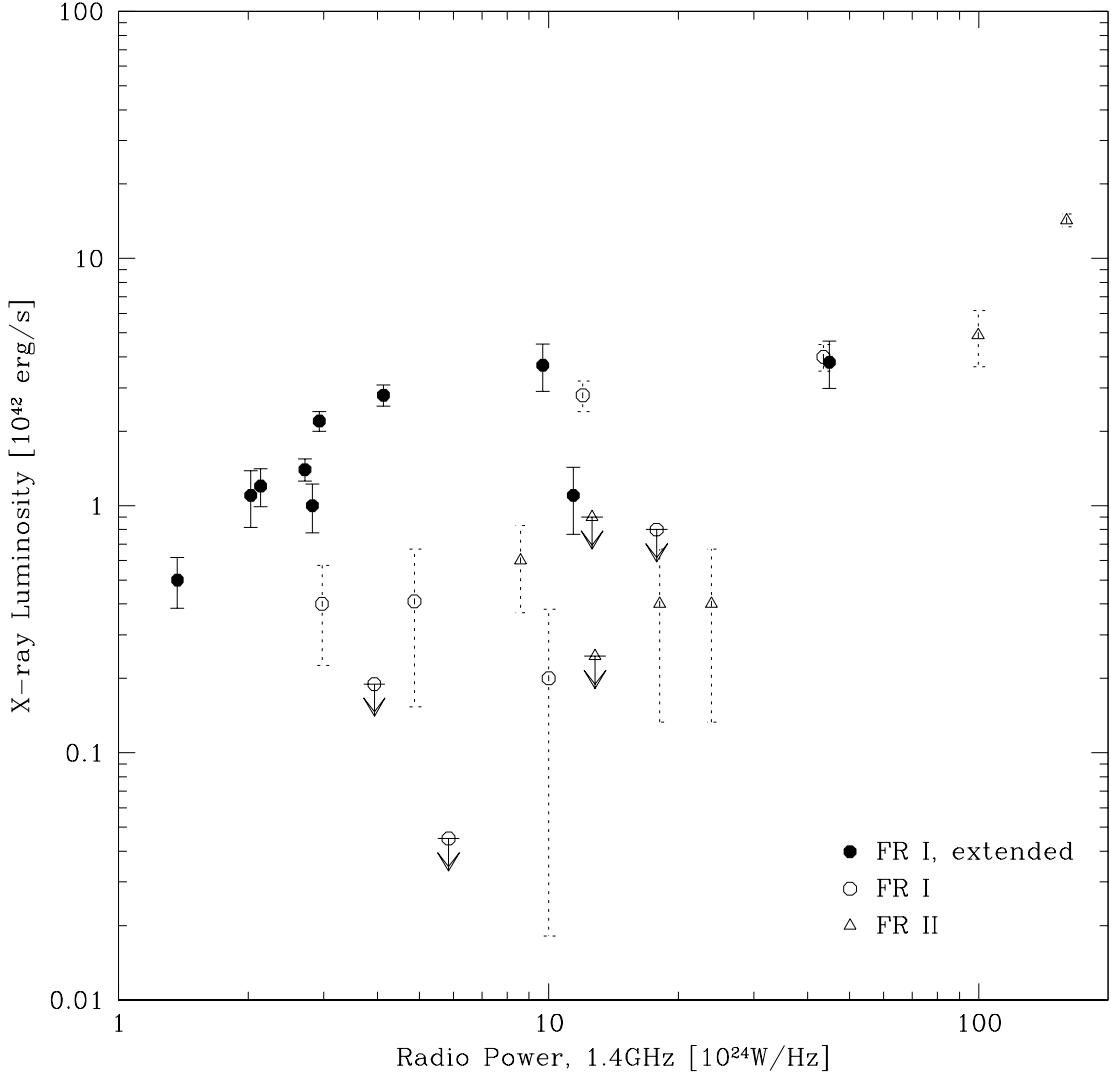}
\caption{X-ray luminosity vs. radio power at 1.4 GHz. The X-ray luminosities are determined from the RASS and represent all flux within a 250 $h_{75}^{-1}kpc$ annulus about the source, and Table 5 lists the references for the radio powers. Filled circles represent FR I sources with extended X-ray emission whereas open circles are FR I sources which are unresolved by RASS. The triangles represent the FR II sources. Upper limits for X-ray luminosities correspond to $2\sigma$, and the error bars are $1\sigma$ and consider only Poissonian counting errors.}
\end{figure}

\begin{figure}
\figurenum{8}
\plotone{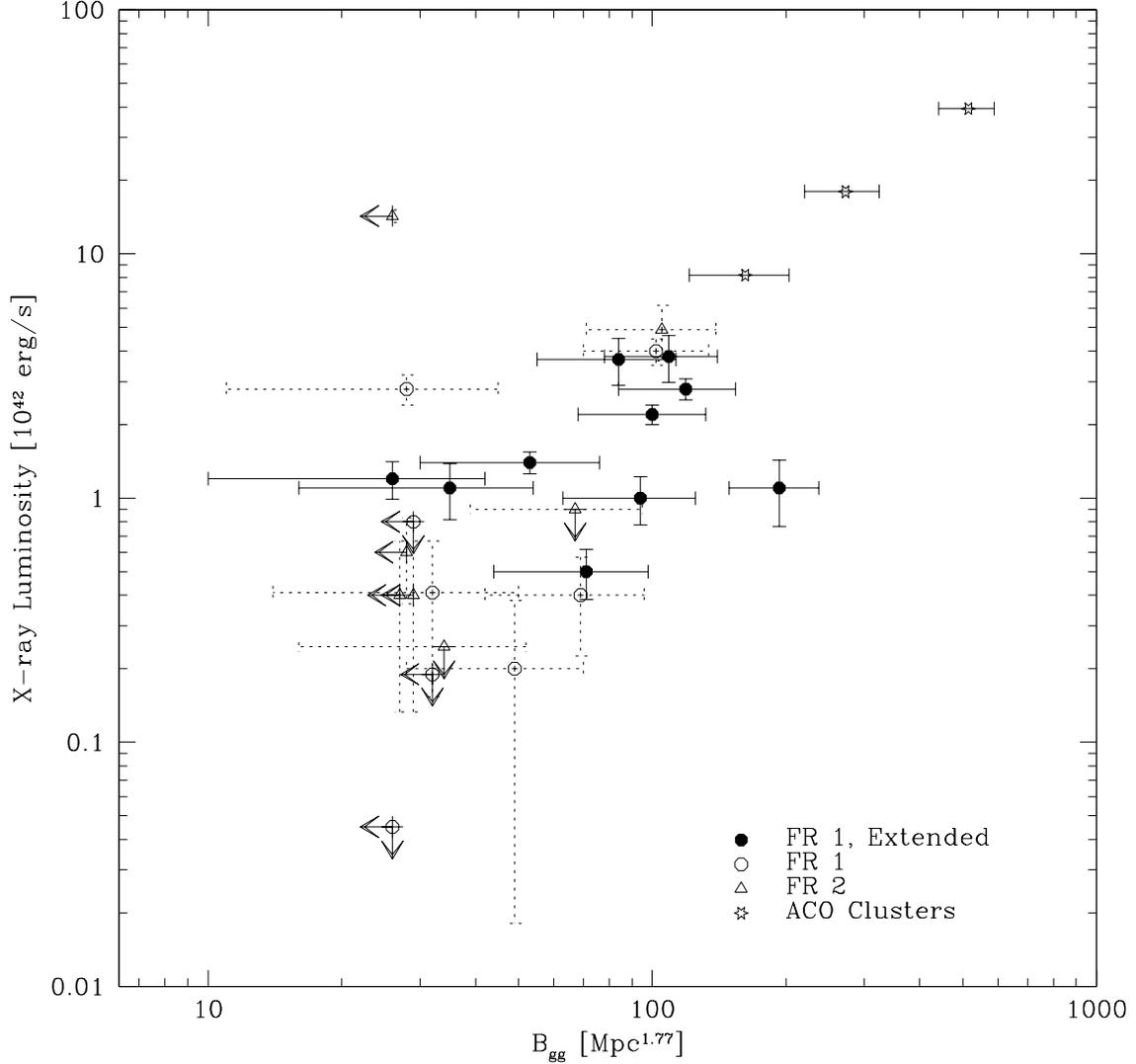}
\caption{$B_{gg}$ vs. X-ray luminosity, where X-ray luminosities are determined from the RASS and represent all flux within a 250 kpc annulus about the source. The method for calculating $B_{gg}$ is explained in the text. Filled circles represent FR I sources with extended X-ray emission whereas open circles are FR I sources which are unresolved by RASS. The triangles represent the FR II sources, and stars represent average values for ACO clusters of richness class 0, 1, and 2 (see Table 3). Error bars are plotted for $1 \sigma$ and are based solely on Poissonian counting errors. The X-ray upper limits correspond to $2\sigma$. Upper limits for $B_{gg}$ represent the value for which the supposed cluster would be significantly above the background at $95\%$ confidence.}
\end{figure}

\begin{table}

\begin{center}
\tablenum{1}
\begin{tabular}{c c c c c}
\multicolumn{5}{c}{Table 1. Sample Information} \\
\hline
\hline
\multicolumn{1}{c}{Source} &
\multicolumn{1}{c}{Class} &
\multicolumn{1}{c}{z} &
\multicolumn{1}{c}{RA} &
\multicolumn{1}{c}{Dec} \cr
\multicolumn{1}{c}{(1)} &
\multicolumn{1}{c}{(2)} &
\multicolumn{1}{c}{(3)} &
\multicolumn{1}{c}{(4)} &
\multicolumn{1}{c}{(5)} \cr
\hline
3C31 & FR I & 0.017 & 01 07 25 & 32 24 45 \\
3C66B & FR I & 0.022 & 02 23 11 & 42 59 32 \\
3C76.1 & FR I & 0.0324 & 03 03 15 & 16 26 19 \\
3C78 & FR I & 0.029 & 03 08 26 & 04 06 39 \\
NGC1692 & FR I & 0.035 & 04 55 24 & -20 34 13 \\
3C278 & FR I & 0.015 & 12 54 37 & -12 33 23 \\
NGC5090 & FR I & 0.011 & 13 21 09 & -43 42 39 \\
3C293 & FR I & 0.045 & 13 49 41 & 31 02 31 \\
3C296 & FR I & 0.024 & 14 16 53 & 10 48 27 \\
3C305 & FR I & 0.0417 & 14 49 21 & 63 16 14 \\
3C310 & FR I/II & 0.054 & 15 04 57 & 26 00 59 \\
NGC6109 & FR I & 0.0296 & 16 17 41 & 35 00 16 \\
NGC6251 & FR I/II & 0.024 & 16 32 32 & 82 32 16 \\
3C386 & FR I & 0.0177 & 18 38 26 & 17 11 49 \\
2152-69 & FR I & 0.027 & 21 57 07 & -69 41 33 \\
3C442A & FR I & 0.0171 & 22 14 47 & 13 50 27 \\
3C449 & FR I & 0.017 & 22 31 21 & 39 21 30 \\
NGC7385 & FR I & 0.0243 & 22 49 55 & 11 36 33 \\
0131-36 & FR II & 0.030 & 01 33 58 & -36 29 36 \\
3C88 & FR II & 0.030 & 03 27 55 & 02 33 44 \\
3C98 & FR II & 0.031 & 03 58 55 & 10 25 47 \\
Pictor A & FR II & 0.035 & 05 19 49 & -45 46 46 \\
DA240 & FR I/II & 0.036 & 07 48 37 & 55 48 59 \\
1637-77 & FR II & 0.043 & 16 44 16 & -77 15 35 \\
3C353 & FR II & 0.030 & 17 20 32 & -00 58 46 \\
\hline
\hline
\end{tabular}

\bigskip
\bigskip
Notes to TABLE 1.

\end{center}

\noindent
(1) Source name; (2) Fanaroff-Riley classification from Owen \& Laing (1989); (3) Redshift; (4) Right ascension of optical center of galaxy, in J2000 coordinates; (5) Declination of optical center of galaxy, in J2000 coordinates.

\end{table}

\begin{table}

\begin{center}
\tablenum{2}
\begin{tabular}{c c c}
\multicolumn{3}{c}{Table 2. Parameters used in $B_{gg}$ Calculation} \\
\hline
\hline
\multicolumn{1}{c}{Parameter} &
\multicolumn{1}{c}{Choice} &
\multicolumn{1}{c}{Reference} \cr
\multicolumn{1}{c}{(1)} &
\multicolumn{1}{c}{(2)} &
\multicolumn{1}{c}{(3)} \cr
\hline
Luminosity Function  & $\phi (\leq M) = \left \{
\matrix{
A 10^{0.75M} & & M < M^* \cr
B 10^{0.25M} & & M \geq M^* \cr
}
\right.
$ &  Abell 1962 \\
 & $\phi (M^*) = 6.3~\times~10^{-3}$ $h_{75}^3Mpc^{-3}$ & Peebles 1971 \\
 & $M_V^*=-20.1$ & \\
Background Galaxy Counts & ${\cal N} = {\cal N}_o 10^{0.6 m_o}$ &  \\
 & ${\cal N}_0 = 1.43~\times~10^{-5}$ ${galaxies \over steradian}$ & Peebles 1973 \\
Limiting magnitude & $m_o$ {\it corresponds to $M_V^*$ at the}& Andersen \& Owen 1994 \\
 &{\it distance of the source} & \\
Angular counting radius & $\theta$ {\it determined such that a cluster with}& Andersen \& Owen 1994 \\
 & $B_{gg}=100$ {\it be detected with S/N=3 }& \\
Power of correlation function & $\gamma = 1.77$ & Groth \& Peebles 1977 \\
\hline
\hline
\end{tabular}
\end{center}

\end{table}

\begin{table}
\begin{center}
\tablenum{3}
\begin{tabular}{c c c c c}
\multicolumn{5}{c}{Table 3. $B_{gg}$ and X-ray Luminosities of Abell Clusters} \\
\hline
\hline
\multicolumn{1}{c}{Abell} &
\multicolumn{1}{c}{$B_{gg}$} &
\multicolumn{1}{c}{$B_{gg}$} &
\multicolumn{1}{c}{$B_{gg}$} &
\multicolumn{1}{c}{$L_x$} \cr
\multicolumn{1}{c}{Richness Class} &
\multicolumn{1}{c}{Andersen \& Owen} &
\multicolumn{1}{c}{Prestage \& Peacock} &
\multicolumn{1}{c}{Yee \& Lopez-Cruz} &
\multicolumn{1}{c}{$10^{42}h_{75}^{-2}~erg~s^{-1}$} \cr
\multicolumn{1}{c}{(1)} &
\multicolumn{1}{c}{(2)} &
\multicolumn{1}{c}{(3)} &
\multicolumn{1}{c}{(4)} &
\multicolumn{1}{c}{(5)} \cr
\hline
0 & 162 $\pm$ 41 & 146 $\pm$ 15 & 293 $\pm$ 98 & 8.2 \\
1 & 272 $\pm$ 52 & 350 $\pm$ 37 & 488 $\pm$ 98 & 18.0 \\
2 & 515 $\pm$ 74 & 500 $\pm$ 61 & 683 $\pm$ 98 & 39.4 \\
\hline
\hline
\end{tabular}

\bigskip
Notes to TABLE 3.
\end{center}

\noindent
(1) Abell's richness class: where 0 indicates 30-49 members, 1 indicates 50-79 members, and 2 indicates 80-109 members; (2) Average $B_{gg}$ per richness class (in units of $h_{75}^{-1.77}Mpc^{1.77}$), from Andersen \& Owen (1994) who use a similar procedure to that adopted in this study; (3) Average $B_{gg}$ as presented by Prestage \& Peacock (1988, 1989), in units of $h_{75}^{-1.77}Mpc^{1.77}$; (4) Suggested $B_{gg}$ per richness class (in units of $h_{75}^{-1.77}Mpc^{1.77}$), from Yee \& Lopez-Cruz (1999). Note that these represent suggested values of $B_{gg}$ by richness class, based on a reasonable linear fit to their results. Their median values of $B_{gg}$ by richness class are roughly 465, 465, and 830; (5) X-ray luminosity, based on the fit of Voges {\it et al.} (1999). This is derived from the RASS using a 500kpc aperature, and the values presented here assume 40 members for an R=0 cluster, 65 for R=1, and 105 for R=2 (i.e., the midpoints for each richness as defined by Abell).
\bigskip

\end{table}

\begin{table}

\begin{center}
\tablenum{4}
\begin{tabular}{c c c c c c}
\multicolumn{6}{c}{Table 4. Model Clusters and Optimal $B_{gg}$ Parameters} \\
\hline
\hline
\multicolumn{1}{c}{Class} &
\multicolumn{1}{c}{$B_{gg}$} &
\multicolumn{1}{c}{$R_c$} &
\multicolumn{1}{c}{z} &
\multicolumn{1}{c}{$m_o$} &
\multicolumn{1}{c}{d} \cr
\multicolumn{1}{c}{(1)} &
\multicolumn{1}{c}{(2)} &
\multicolumn{1}{c}{(3)} &
\multicolumn{1}{c}{(4)} &
\multicolumn{1}{c}{(5)} &
\multicolumn{1}{c}{(6)} \cr
\hline
RG & 73 & 50 & 0.03 & $m^*+1.0$ & 0.9 \\
ACO 0 & 289 & 100 & 0.07 & $m^*+1.0$ & 1.3 \\
ACO 1 & 487 & 100 & 0.07 & $m^*+1.1$ & 1.9 \\
ACO 2 & 694 & 100 & 0.07 & $m^*+1.2$ & 2.1 \\
\hline
\hline
\end{tabular}
\end{center}

\begin{center}
Notes to TABLE 4.
\end{center}

\noindent
(1) Model representation, where `RG' represents a typical radio galaxy from this study and `ACO' represent Abell clusters of richness class 0, 1, and 2; (2) $B_{gg}$ value, corresponding to parameters of Yee \& Lopez-Cruz but with $H_o=75$ km/sec/Mpc; (3) Core radius, in kpc, used in the King profile of the model; (4) Redshift of the simulated cluster; (5) Limiting magnitude to count down to in order to maximize S/N for each cluster model; (6) Linear counting radius, in Mpc, in order to maximize SN for each cluster model.
\bigskip

\end{table}

\begin{table}

\begin{center}
\tablenum{6}
\begin{tabular}{c c c c c c c c}
\multicolumn{8}{c}{Table 6. Correlations of Radio Power, X-Ray Luminosity, and $B_{gg}$} \\
\hline
\hline
\multicolumn{2}{c}{ } &
\multicolumn{2}{c}{$P_{1.4GHz}-L_x$} &
\multicolumn{2}{c}{$P_{1.4GHz}-B_{gg}$} &
\multicolumn{2}{c}{$L_x-B_{gg}$} \cr
\multicolumn{1}{c}{Sample} &
\multicolumn{1}{c}{Number of Points} &
\multicolumn{1}{c}{$\tau$} &
\multicolumn{1}{c}{P} &
\multicolumn{1}{c}{$\tau$} &
\multicolumn{1}{c}{P} &
\multicolumn{1}{c}{$\tau$} &
\multicolumn{1}{c}{P} \cr
\multicolumn{1}{c}{(1)} &
\multicolumn{1}{c}{(2)} &
\multicolumn{1}{c}{(3)} &
\multicolumn{1}{c}{(4)} &
\multicolumn{1}{c}{(5)} &
\multicolumn{1}{c}{(6)} &
\multicolumn{1}{c}{(7)} &
\multicolumn{1}{c}{(8)} \cr
\hline
All FR I and FR II & 25 & 0.280 & 0.700 & -0.067 & 0.193 & 0.621 & 0.982\\
FR I & 18 & 0.366 & 0.722 & 0.235 & 0.519 & 0.797 & 0.985 \\
FR II & 7 & 1.048 & 0.964 & -0.095 & 0.133 & -0.095 & 0.149 \\
FR I, extended & 10 & 1.244 & 0.988 & 1.200 & 0.984 & 0.444 & 0.631 \\
\hline
\hline
\end{tabular}
\bigskip

\bigskip
Notes to TABLE 6.
\end{center}

\noindent
(1) Types of sources included in test; (2) Number of sources included in test; (3) Kendall's $\tau$ for correlation of radio power at 1.4 GHz with RASS-derived x-ray luminosity; (4) Corresponding probability that radio power and x-ray luminosity are correlated;  (5) Kendall's $\tau$ for correlation of radio power at 1.4 GHz with $B_{gg}$; (6) Corresponding probability that radio power and $B_{gg}$ are correlated; (7) Kendall's $\tau$ for correlation of RASS-derived x-ray luminosity with $B_{gg}$; (8) Corresponding probability that x-ray luminosity and $B_{gg}$ are correlated.
\bigskip

\end{table}

\end{document}